\documentclass[11pt]{article}
\usepackage[left=0.6in,right=0.6in]{geometry}
\usepackage{empheq}
\usepackage{amssymb}
\usepackage{dsfont}
\usepackage{slashed}
\usepackage[pdfstartview=FitH,pdfpagemode=None]{hyperref}
\usepackage{color}
\usepackage{multirow}
\usepackage{mathtools}
\usepackage{stackrel}
\usepackage{enumerate}
\usepackage{bm}

\newcommand{\be}{\begin{equation}}
\newcommand{\ee}{\end{equation}}
\newcommand{\ba}{\begin{eqnarray}}
\newcommand{\ea}{\end{eqnarray}}
\newcommand{\bea}{\begin{eqnarray}}
\newcommand{\eea}{\end{eqnarray}}

\makeatletter\@addtoreset{equation}{section}\makeatother

\begin{document}

\begin{titlepage}
\hfill LCTP-19-13

\begin{center}
\vspace{1cm}

{\huge \textbf{ One-loop Holography with  Strings in $AdS_4\times\mathbb {CP}^3$}}\\[4em]

{\small Marina David${}^1$, Rodrigo de Le\'on Ard\'on ${}^{2}$,  Alberto  Faraggi$^{3}$, Leopoldo A.~Pando Zayas${}^{1,2}$ and Guillermo A. Silva${}^{4}$}\\[3em]
%
%

${}^{1}$\emph{Leinweber Center for Theoretical Physics,  Randall Laboratory of Physics\\ The University of
Michigan,  Ann Arbor, MI 48109, USA}\\[1em]

${}^2$\emph{The Abdus Salam International Centre for Theoretical Physics\\ Strada Costiera 11,  34014 Trieste, Italy}\\[1em]

${}^{3}$\emph{Departamento de Ciencias F\'isicas,  Facultad de Ciencias Exactas\\
Universidad Andr\'es Bello,  Sazie 2212, Piso 7, Santiago, Chile}\\[1em]

${}^{4}$\emph{Instituto de F\'isica de La Plata - CONICET \&\\ Departamento de F\'isica, 
 UNLP C.C. 67, 1900 La Plata, Argentina}\\[3em]

\abstract{We compute the one-loop effective action of string configurations embedded in $AdS_4\times\mathbb{CP}^3$ which are dual to $\frac{1}{6}$-BPS latitude Wilson Loops in the ABJM theory. To avoid ambiguities in the string path integral we subtract the $\frac{1}{2}$-BPS case. The one-loop determinants are computed by Fourier-decomposing the two dimensional operators and then using the Gel'fand-Yaglom method. We comment on various aspects related to the regularization procedure, showing the cancellation of a hierarchy of divergences. After taking into account an IR anomaly from a change in topology, we find a precise agreement with the field theory result known from supersymmetric localization.

%

}

\vspace{1cm}
{ \tt  mmdavid@umich.edu, rde\_leon@ictp.it, alberto.faraggi@unab.cl,}\\
{\tt lpandoz@umich.edu, silva@fisica.unlp.edu.ar}
\end{center}

\end{titlepage}

\tableofcontents

\section{Introduction}

Supersymmetric localization has provided a wealth of exact results in many supersymmetric field theories  including various with known gravity duals.  The original work of Pestun \cite{Pestun:2007rz}, addressing Wilson loops in ${\cal N}=4 $ supersymmetric Yang-Mills in four dimensions, prompted  holographic computations beyond the leading order stimulating much activity over the past ten years \cite{Drukker:2000ep, Kruczenski:2008zk, Forini:2015bgo, Faraggi:2016ekd, Aguilera-Damia:2018twq, Cagnazzo:2017sny, Medina-Rincon:2018wjs}.

Following the original large $N$ evaluation of Wilson loops in the ABJM theory  in \cite{Kapustin:2009kz}, a number of results have provided  answers that are exact in various parameters up to exponentially small corrections. For example,  the exact expectation value of the $\frac{1}{2}$-BPS  was obtained in \cite{Klemm:2012ii} (see also \cite{Okuyama:2016deu}) and, more recently, for the $\frac{1}{6}$-BPS configuration corresponding to the latitude Wilson loop the exact expectation value was obtained in  \cite{Bianchi:2018bke}. 

It is natural to turn the tools of precision holography to this setup and, indeed,  a subset of the authors addressed this problem in \cite{Aguilera-Damia:2018bam} using the zeta-function regularization tools developed in \cite{Aguilera-Damia:2018rjb}.   In this manuscript we report on the one-loop effective action of the corresponding strings using the method of Gel'fand-Yaglom. Our result perfectly matches the field theory result in the appropriate regime of parameters corresponding to the supergravity regime for large $N$ and large `t Hooft coupling $\lambda$. 

The goal of precision holography has been to use the results of field theory to sharpen and develop new  tools  to tackle the supergravity side beyond the leading order.  It is fair to say that this program is now not only bearing fruits but also shedding some light on the details of various technical methods and clarifying the structure of semi-classical string perturbation theory on curved backgrounds with Ramond-Ramond fluxes.  We hope that some of the lessons learned in this arena will be valuable to other precision holography endeavours such as the computation of quantum corrections to the entropy of black holes whose field theory duals are by now well understood. 

In  section \ref{Sec:FieldClass} we briefly discuss the field theory result that we aim to match as well as the defining properties of the classical string configurations. Section \ref{Sec:1loop}  contains an abridged presentation of the spectrum of fluctuations (see  \cite{Aguilera-Damia:2018bam} for more explicit details) and a summary of the computations of one-loop determinants. We discuss the explicit details of the cancellation of potential logarithmic divergences and comment on fermionic boundary conditions in section \ref{Sec:discussion}. We conclude in section \ref{Sec:Conclusions} with a summary of our work and point out some interesting open questions in precision holography with Wilson loops. \\

{\bf Note}: While preparing this manuscript for publication, a precise match with the field theory prediction was reported in \cite{Medina-Rincon:2019bcc} using the method of phase shifts. Although there is considerable overlap, our methods differ substantially.

\section{The latitude $\frac{1}{6}$-BPS Wilson loop and its holographic dual}\label{Sec:FieldClass}

The exact expectation value of the so-called fermionic $\frac{1}{6}$-BPS latitude Wilson loop in ABJM is given by \cite{Bianchi:2018bke}
\bea
\label{Eq:FieldTheory}
\langle W_F^{\frac{1}{6}} (\nu) \rangle  =  \frac{i  \nu \Gamma \left(- \frac{\nu}{2} \right) \sin \left( \frac{\pi \nu}{2} \right) \mathrm{Ai} \left[ \left(\frac{2}{\pi^2 k} \right)^{-1/3} \left( N - \frac{k}{24} - \frac{6 \nu + 1}{3k} \right)\right]}{2^{\nu+1} \sqrt{\pi} \Gamma \left( \frac{3-\nu}{2}\right) \sin \left( \frac{2 \pi \nu}{k}\right) \mathrm{Ai} \left[ \left( \frac{2}{\pi^2 k} \right)^{-1/3} \left( N - \frac{k}{24} - \frac{1}{3k}\right)\right]},
\eea
where $\nu = \cos \theta_0$ determines the latitude angle. Notice that this is the un-normalized version of the Wilson loop. This result was preceded by an impressive series of papers \cite{Cardinali:2012ru,Bianchi:2014laa,Bonini:2016fnc,Bianchi:2017svd,Bianchi:2017ozk,Bianchi:2018scb}. 

When expanded in the holographic regime, namely, taking the genus-zero contribution at leading order in $\lambda=N/k$, \eqref{Eq:FieldTheory} coincides with the minimal area of the dual $\frac{1}{6}$-BPS string on $AdS_4\times\mathds{CP}^3$ \cite{Correa:2014aga}, whose induced metric on the world-sheet reads
\begin{empheq}{alignat=7}\label{induced}
	ds^2_{\textrm{ind}}&=A^2\,ds^2_{\textrm{cyl}}\,,
	&\qquad
	A^2&=\sinh^2\rho+\sin^2\vartheta_1\,,
	&\qquad
	ds^2_{\textrm{cyl}}&=d\sigma^2+d\tau^2\,,
\end{empheq}
where,
\begin{empheq}{alignat=7}
	\sinh\rho&=\frac{1}{\sinh\sigma}\,,
	&\qquad
	\sin\vartheta_1&=\frac{1}{\cosh\left(\sigma+\sigma_0\right)}\,,
	&\qquad
	\sigma&>0\,,
	&\qquad
	\tau&\sim\tau+2\pi\,.
\end{empheq}
The integration constant $\sigma_0$ is related to the latitude angle of the Wilson loop via $\cos\theta_0=\tanh \sigma_0$.

The goal of this manuscript is to check that there is still agreement between the gauge theory and the gravity dual after including the first quantum corrections in $1/\sqrt{\lambda}$. In order to avoid subtleties in the string path integral measure, we consider the ratio between the $\frac{1}{6}$-BPS Wilson loop expectation value and its $\frac{1}{2}$-BPS limit. The latter corresponds to $\theta_0\rightarrow0$ ($\nu\to1$, $\sigma_0\to\infty$), for which the induced geometry becomes exactly $AdS_2$. Thus, expanding \eqref{Eq:FieldTheory} in the appropriate limit, we find that the field theory prediction to be matched with the one-loop effective action of the string configuration takes the form
\bea\begin{split}\label{WLth}
 \Delta \Gamma^{\text{1-loop}}_{\text{effective}} = \text{ln} \left[ \frac{\langle W_F^{\frac{1}{6}} (1) \rangle}{\langle W_F^{\frac{1}{6}} (\nu) \rangle} \right] &=  -\ln \Gamma \left( \frac{\nu + 1}{2}\right) + \ln \Gamma ( \nu + 1) + \ln \Gamma \left( \frac{3}{2} - \frac{\nu}{2}\right). \\
\end{split}\eea


\section{One-loop effective action }\label{Sec:1loop}
On the holographic side, the difference in one-loop effective actions between the $\frac{1}{6}$-BPS and the $\frac{1}{2}$-BPS strings is given by
\bea\label{eq6}
e^{-\Delta\Gamma^{\text{1-loop}}_{\text{effective}} (\theta_0)} = 
\left[\frac{ {\displaystyle \prod_{\alpha=\pm}   \frac{\text{det}\: \mathcal{O}^{1,\alpha} (\theta_0)}{\text{det}\: \mathcal{O} ^{1,\alpha} (0)}  \: 
\left( \frac{\text{det}\: \mathcal{O}^{2,\alpha}  (\theta_0)}{\text{det}\: \mathcal{O}^{2,\alpha}  (0)} \right)^2  \: 
 \frac{\text{det}\: \mathcal{O}^{3,\alpha} (\theta_0)}{\text{det}\: \mathcal{O}^{3,\alpha} (0)}  }}
{  {\displaystyle\Big{(} \frac{\text{det}\: \mathcal{O}^4 (\theta_0)}{\text{det}\: \mathcal{O}^4 (0)}\Big{)}^2\:  
\prod_{\alpha=\pm} \frac{\text{det}\: \mathcal{O}^{5,\alpha}   (\theta_0)}{\text{det}\: \mathcal{O}^{5,\alpha}   (0)}  \:
\left(  \frac{\text{det}\: \mathcal{O}^{6,\alpha} (\theta_0)}{\text{det}\: \mathcal{O}^{6,\alpha} (0)}\right)^2}  }\right]^{\frac{1}{2}}.
\eea
The precise form of the operators, computed originally in \cite{Aguilera-Damia:2018bam}, is spelled out in section \ref{SQF}. Roughly speaking, after freezing the longitudinal modes, $\mathcal{O}^4$ comes from the two normal fluctuations of the string in $AdS_4$, whereas $\mathcal{O}^{5,\alpha}$ and $\mathcal{O}^{6,\alpha}$ correspond to the six fluctuations in $\mathds{CP}^3$. The fermionic fields give rise to $\mathcal{O}^{1,\alpha}$, $\mathcal{O}^{2,\alpha}$ and $\mathcal{O}^{3,\alpha}$.


The determinants in \eqref{eq6} are defined using the string induced metric \eqref{induced}, which has the topology of a disk. It is convenient, however, to strip away the conformal factor $A$ and compute all quantities using the cylinder metric $ds^2_{\textrm{cyl}}$. This transformation has two important effects: 
\begin{enumerate}[i)]
	\item there is a potential Weyl anomaly in the determinants due to the rescaling of the metric,
	\item there is an additional IR anomaly due to the change in topology of the string worldsheet.
\end{enumerate}
Since passing to the cylinder corresponds to choosing a conformal gauge, the Weyl anomaly actually vanishes, as it should in critical String Theory. The second effect was discussed in \cite{Cagnazzo:2017sny}, and amounts to correcting the 1-loop effective action by
\begin{empheq}{alignat=7}\label{IR anomaly}
	\Gamma^{\text{1-loop}}_{\text{effective}}&\longrightarrow\Gamma^{\text{1-loop}}_{\text{effective}}+\Gamma_{\infty}\,,
	&\qquad
	\Gamma_{\infty}&=\frac{1}{2}\ln\left(\frac{1+\cos\theta_0}{2}\right)\,.
\end{empheq}
This correction, whose origin can be explained by the use of a diffeomorphic-invariant regulator, must be taken into account in order to get a precise match with the field theory prediction. In what follows, all quantities will be defined with respect to the flat cylinder metric.

\subsection{Spectrum}
\label{SQF}
After stripping away the conformal factor, the bosonic operators in the spectrum of fluctuations of the $\frac{1}{6}$-BPS string are \cite{Aguilera-Damia:2018bam}
\begin{empheq}{alignat=7}\label{bosonic operators}
	\mathcal{O}^4 &=-\partial_{\sigma}^2-\partial_{\tau}^2+2\sinh^2\rho\,,
	\\
	\mathcal{O}^{5,\alpha}&=-\partial_{\sigma}^2-\left(\partial_{\tau}+i\alpha\mathcal{A}\right)^2-\partial_{\sigma}\mathcal{A}\,,
	\\
	\mathcal{O}^{6,\alpha}&=-\partial_{\sigma}^2-\left(\partial_{\tau}-i\alpha\mathcal{B}\right)^2+\partial_{\sigma}\mathcal{B}\,,
\end{empheq}
where
\begin{empheq}{alignat=7}
	\mathcal{A}&=\left(\frac{\cosh\rho\cos\vartheta_1+1}{\cosh\rho+\cos\vartheta_1}-1\right)\,,
	&\qquad
	\mathcal{B}&=\frac{1}{2}\left(1-\cos\vartheta_1\right)\,.
\end{empheq}
Similarly, upon reducing the type IIA spinors down to two dimensions, the fermionic operators read \cite{Aguilera-Damia:2018bam}
\begin{empheq}{alignat=7}
	\mathcal{O}^{\alpha,\beta,\gamma}&=-i\gamma_1\partial_{\sigma}-i\gamma_0\left(\partial_{\tau}+\frac{i\alpha}{2}\mathcal{A}+\frac{i(\beta+\gamma)}{2}\mathcal{B}\right)-M^{\alpha,\beta,\gamma}\,,
\end{empheq}
with
\begin{empheq}{alignat=7}
	M^{\alpha,\beta,\gamma}&=\frac{1}{4}\left(3\beta\gamma-1\right)\left(\alpha\sinh^2\rho\gamma^*+\sin^2\vartheta_1\right)A^{-1}+\frac{1}{4}(\beta+\gamma)A\gamma^*\,.
\end{empheq}
The labels $\alpha$, $\beta$ and $\gamma$ take values $\pm1$ and $\gamma^*=-i\gamma_{01}$ is the chirality matrix. The operators appearing in the 1-loop effective action are defined as
\begin{empheq}{alignat=7}\label{fermionic operators}
	\mathcal{O}^{1,\alpha}&=\mathcal{O}^{\alpha,\alpha,\alpha}\,,
	&\qquad
	\mathcal{O}^{2,\alpha}&=\mathcal{O}^{\alpha,\beta,-\beta}\,,
	&\qquad
	\mathcal{O}^{3,\alpha}&=\mathcal{O}^{\alpha,-\alpha,-\alpha}\,.
\end{empheq}
Notice that $\mathcal{O}^{\alpha,\beta,-\beta}$ does not depend on the label $\beta$. From now on we work in the representation
\begin{empheq}{alignat=7}\label{Dirac basis}
	\gamma_0&=\sigma_2\,,
	&\qquad
	\gamma_1&=\sigma_1\,,
	&\qquad\Rightarrow\qquad
	\gamma^*&=-\sigma_3\,.
\end{empheq}
For both bosons and fermions, we refer to the operators $\mathcal{O}^{\alpha}$ and $\mathcal{O}^{-\alpha}$ as charge conjugates of each other.
\subsection{Determinants and boundary conditions}
By now, a considerable body of work exists showing the merits and drawbacks of the different techniques used to compute functional determinants. In this manuscript, we will take advantage of the rotational symmetry of the worldsheet and Fourier-decompose the two-dimensional operators into an infinite number of one-dimensional ones,
\begin{empheq}{alignat=7}
	\mathcal{O}_E&=\mathcal{O}\Big|_{\partial_{\tau}\to-iE}\,,
\end{empheq}
with $E\,\in\,\mathds{Z}$ for bosons and $E\,\in\,\mathds{Z}+\frac{1}{2}$ for fermions. Then, we will use the Gel'fand-Yaglom method \cite{Gelfand:1959nq,Forman1987} to compute the corresponding ratio of determinants along the radial direction and sum over the Fourier modes. This procedure has been applied to a number of problems in the context of holographic Wilson loops and we refer the interested reader to \cite{Kruczenski:2008zk,Forini:2015bgo,Faraggi:2016ekd,Kristjansen:2012nz} for details.

As usual, functional determinants suffer from divergences that demand a careful treatment. In the present context, the infinite volume of the worldsheet requires the introduction of a UV cutoff for the radial coordinate at $\sigma=\epsilon$, as well as a large IR regulator at $\sigma=R$. As it turns out, these subtleties are taken care of by considering the ratio of determinants between the $\frac{1}{6}$-BPS and $\frac{1}{2}$-BPS configurations, which renders the $\epsilon\to0$ and $R\to\infty$ limits well-defined. Additionally, there are divergences coming from the sum over Fourier modes which require an additional UV cutoff $\Lambda$. Even though these divergences are unavoidable for each individual determinant, even after taking the ratio, they end up canceling due to supersymmetry and the intricacies of the string spectrum of fluctuations. We will say more about this in the discussion section.

A key ingredient in computing any determinant are the boundary conditions imposed on the fields. The treatment of bosonic boundary conditions is standard so we avoid presenting too many details. We follow the same procedure as in \cite{Forini:2015bgo,Faraggi:2016ekd}. It suffices to recall that, according to the Gel'fand-Yaglom method, the ratio of determinants between the $\frac{1}{6}$-BPS and the $\frac{1}{2}$-BPS radial operators with Dirichlet-Dirichlet boundary conditions on the interval $[\epsilon,R]$ is given by
\begin{empheq}{alignat=7}\label{GY bosons}
	\Omega_E(\theta_0)&\equiv\frac{\det\mathcal{O}_E(\theta_0)}{\det\mathcal{O}_E(0)}=\lim_{\epsilon\to0}\lim_{R\to\infty}\frac{\chi(R)}{{\displaystyle\lim_{\theta_0\to0}\chi(R)}}\,,
\end{empheq}
where $\chi$ is the solution to the intial value problem
\begin{empheq}{alignat=7}\label{GY eq bosons}
	\mathcal{O}_E\chi&=0\,,
	&\qquad
	\chi(\epsilon)&=0\,,
	&\qquad
	\chi'(\epsilon)&=1\,.
\end{empheq}
As mentioned above, the $\epsilon\to0$ and $R\to\infty$ limits can be safely taken in this ratio. Although the prescription varies for other choices of boundary conditions, in the present problem all choices give the same result.

The real issue lies in the fermionic sector, where the Gel'fand-Yaglom method is slightly more involved and the choice of boundary conditions does affect the final result. In this case, one must first solve the (first order, two-component) differential equation $\mathcal{O}_E=0$. The two linearly independent solutions can be conveniently organized into a $2\times2$ matrix $Y(\sigma)$ satisfying
\begin{empheq}{alignat=7}
	\mathcal{O}_EY&=0\,,
	&\qquad
	Y(\epsilon)&=\mathds{1}_{2\times2}\,,
	&\qquad
	Y(\sigma)&=\left(
	\begin{array}{cc}
		\psi_1^{I}(\sigma) & \psi_1^{II}(\sigma)
		\\
		\psi_2^{I}(\sigma) & \psi_2^{II}(\sigma)
	\end{array}
	\right)\,.
\end{empheq}
Here, the subscript labels the two components of the spinor and the superscript the two independent solutions. The ratio of determinants is then given by the expression\footnote{For operators of the form $\mathcal{O}=P_0\partial_{\sigma}+P_1$, the determinants include a prefactor involving $\textrm{Tr}\left(\mathcal{R}P_1P_0^{-1}\right)$, where $\mathcal{R}$ is a projector selecting half of the eigenvalues of $P_0$. To avoid this subtlety, we compute instead the determinants of $P_0^{-1}\mathcal{O}=\partial_{\sigma}+P_0^{-1}P_1$. Then, according to lemma 3.1 in \cite{Forman1987}, one can choose $\mathcal{R}=0$.}
\begin{empheq}{alignat=7}\label{GY fermions}
	\Omega_E(\theta_0)&\equiv\frac{\det\mathcal{O}_E(\theta_0)}{\det\mathcal{O}_E(0)}=\lim_{\epsilon\to0}\lim_{R\to\infty}\frac{\det\left(M+NY(R)\right)}{{\displaystyle\lim_{\theta_0\to0}}\det\left(M+NY(R)\right)}\,,
\end{empheq}
where the $2\times 2$ matrices $M$ and $N$ parametrize the boundary conditions at $\sigma=\epsilon$ and $\sigma=R$ via 
\begin{empheq}{alignat=7}
	M\psi(\epsilon)+N\psi(R)&=0\,.
\end{empheq}
In the present case, the fermionic operators \eqref{fermionic operators} satisfy $\mathcal{O}^{-\alpha}_E=\gamma_1\mathcal{O}^{\alpha}_{-E}\gamma_1$. Since we expect charge conjugate fields to contribute identically to the one-loop effective action, this motivates relating the boundary conditions for $\mathcal{O}^{\alpha}_E$ and $\mathcal{O}^{-\alpha}_E$ by
\begin{empheq}{alignat=7}\label{conjugate bc}
	M^{-\alpha}&=\gamma_1M^{\alpha}\gamma_1\,,
	&\qquad
	N^{-\alpha}&=\gamma_1N^{\alpha}\gamma_1\,.
\end{empheq}
We then choose
\begin{empheq}{alignat=7}\label{Matrix M}
	M^+&=\left(
	\begin{array}{cc}
		M_1 & M_2
		\\
		0 & 0
	\end{array}
	\right)\,,
	&\qquad
	N^+&=\left(
	\begin{array}{cc}
		0 & 0
		\\
		N_1 & N_2
	\end{array}
	\right)\,,
	\quad M_i,N_j\in\mathbb{C},
\end{empheq}
so as not to mix the conditions at the two boundaries (i.e. to have local boundary conditions). As we will see below, the boundary conditions at $\sigma=R$ are irrelevant, but at $\sigma=\epsilon$ only a particular choice of $M_1$ and $M_2$ gives the correct answer.
\subsubsection{Bosons}
We now compute the ratio of bosonic determinants with Dirichlet-Dirichlet boundary conditions in $\sigma\,\in\,[\epsilon,R]$ using \eqref{GY bosons}. For the operator $\mathcal{O}_E^4$ this trivial since it does not depend on $\theta_0$. Thus,
\begin{empheq}{alignat=7}\label{Om4}
	\ln\Omega_{E}^{4}&=0\,.
\end{empheq}
For $\mathcal{O}_E^{5,\alpha}$ and $\mathcal{O}_E^{6,\alpha}$, the general solution to the equation $\mathcal{O}_E\chi=0$ reads \cite{Forini:2015bgo,Faraggi:2016ekd}
\begin{empheq}{alignat=7}
	\chi&=e^{-\mathcal{W}}\left(C_1+C_2\int d\sigma\,e^{2\mathcal{W}}\right)\,,
\end{empheq}
with
\begin{empheq}{alignat=7}
	\mathcal{W}^5&=-\alpha E\sigma+\int d\sigma\,\mathcal{A}\,,
	&\qquad
	\mathcal{W}^6&=-\alpha E\sigma-\int d\sigma\,\mathcal{B}\,.
\end{empheq}
Imposing the boundary conditions \eqref{GY eq bosons} and taking the ratios we find
~\\
\begin{empheq}{alignat=7}
	\label{O5}
	\Omega_E^{5,\alpha }&=\left\{
	\begin{array}{cc}
		{\displaystyle\left(\frac{1+\cos\theta_0}{2}\right)^{\frac{1}{2}}} & \alpha E\leq0
		\\\\
		{\displaystyle\left(\frac{1+\cos\theta_0}{2}\right)^{-\frac{1}{2}}\left(\frac{\alpha E+1+\cos\theta_0}{\alpha E+2}\right)} & \alpha E\geq0
	\end{array}
	\right.\,,
\end{empheq}
~\\
\begin{empheq}{alignat=7}
	\label{O6}
	\Omega_E^{6,\alpha }&=\left\{
	\begin{array}{cc}
		{\displaystyle\left(\frac{1+\cos\theta_0}{2}\right)^{\frac{1}{2}}} & \alpha E\leq0
		\\\\
		{\displaystyle\left(\frac{1+\cos\theta_0}{2}\right)^{-\frac{1}{2}}\left(\frac{\alpha E+\frac{1+\cos\theta_0}{2}}{\alpha E+1}\right)} & \alpha E\geq0
	\end{array}
	\right.\,.
\end{empheq}
~\\
Notice that both regulators, $\epsilon$ and $R$, have disappeared. It is convenient at this point to combine the charge conjugate operators into a single expression, namely,\footnote{Even though the operator $\mathcal{O}^4$ is real, we define $\bm\Omega^4$ in this way for notational convenience.}
\begin{empheq}{alignat=7}
	\ln\bm\Omega^4_E&\equiv\ln\Omega_{E}^4=0\,,\label{Om4}
	\\
	\ln\bm\Omega^5_E&\equiv\frac{1}{2}\left(\ln\Omega_{E}^{5,+}+\ln \Omega_{E}^{5,-}\right)&&=\frac{1}{2}\ln\left(\frac{|E|+1+\cos\theta_0}{|E|+2}\right)\,,\label{Om5}
	\\
	\ln\bm\Omega^6_E&\equiv\frac{1}{2}\left(\ln\Omega_{E}^{6,+}+\ln \Omega_{E}^{6,-}\right)&&=\frac{1}{2}\ln\left(\frac{|E|+\frac{1+\cos\theta_0}{2}}{|E|+1}\right)\,.\label{Om6}
\end{empheq}
This has the effect of removing the $E$-independent prefactors in $\Omega_E$, making the absence of linear $\Lambda$ divergences manifest.

\subsubsection{Fermions}
Let us move on to the fermionic fields. After Fourier-transforming $\partial_{\tau}\to-iE$ and defining the spinor projections
\begin{empheq}{alignat=7}
	\psi_{\pm}&\equiv\frac{1}{2}\left(1\mp\alpha\gamma^*\right)\psi&&=\frac{1}{2}\left(
	\begin{array}{c}
		(1\pm\alpha)\psi_1
		\\
		(1\mp\alpha)\psi_2
	\end{array}
	\right)\,,
\end{empheq}
the equation $\mathcal{O}_E\psi=0$ can be written as
\begin{empheq}{alignat=7}
	-i\gamma_1D^{\mp}_{\sigma}\psi_{\pm}-M_{\pm}\psi_{\mp}&=0\,,	
\end{empheq}
with
\begin{empheq}{alignat=7}
	D_{\sigma}^{\pm}&=\partial_{\sigma}\pm\left(-\alpha E+\frac{1}{2}\mathcal{A}+\frac{\alpha(\beta+\gamma)}{2}\mathcal{B}\right)\,,
	\\
	M_{\pm}&=\pm\frac{1}{4}\left(3\beta\gamma-1\right)\left(\sinh^2\rho\pm\sin^2\vartheta_1\right)A^{-1}\pm\frac{1}{4}\alpha(\beta+\gamma)A\,.
\end{empheq}
Notice that the projections $\psi_{\pm}$ depend on the charge $\alpha$. The general solution for the operators $\mathcal{O}^{1,\alpha}$ and $\mathcal{O}^{2,\alpha}$, corresponding, respectively, to $\beta=\gamma=\alpha$ and $\beta=-\gamma$, is given by \cite{Forini:2015bgo,Faraggi:2016ekd}
\begin{empheq}{alignat=7}
	\psi_+&=A^{\frac{1}{2}}e^{-\mathcal{W}}\left(C_1+C_2\int d\sigma\,e^{2\mathcal{W}}\right)\,,
	\\
	\psi_-&=-ip\,A^{-\frac{1}{2}}\gamma_1\left[-2e^{-\mathcal{W}}\left(C_1+C_2\int d\sigma\,e^{2\mathcal{W}}\right)\partial_{\sigma}\mathcal{W}+C_2e^{\mathcal{W}}\right]\,,
\end{empheq}
where $p=\left(3\beta\gamma-1+\alpha(\beta+\gamma)\right)/4=\pm1$. Similarly, for $\mathcal{O}^{3,\alpha}$ ($\beta=\gamma=-\alpha$) one has
\begin{empheq}{alignat=7}
	\psi_+&=A^{\frac{1}{2}}e^{\mathcal{W}}C_+\,,
	\\
	\psi_-&=A^{-\frac{1}{2}}e^{-\mathcal{W}}\left(C_-+i\gamma_1C_+\int d\sigma\,\sin^2\vartheta_1e^{2\mathcal{W}}\right)\,.
\end{empheq}
These last modes become massless in the $\frac{1}{2}$-BPS limit $\theta_0=0$, which explains the decoupling between $\psi_+$ and $\psi_-$. In all cases the prepotential is
\begin{empheq}{alignat=7}
	\mathcal{W}&=-\left(\alpha E+\frac{1}{2}-\frac{1}{4}\alpha(\beta+\gamma)\right)\sigma-\frac{1}{2}\ln\sinh\rho-\frac{1}{4}\left(2-\alpha(\beta+\gamma)\right)\ln\sin\vartheta_1\,.
\end{empheq}

Using the above solutions one can construct the fundamental matrix $Y$ and compute the determinants entering in the Gel'fand-Yaglom formula \eqref{GY fermions} with boundary conditions \eqref{conjugate bc}-\eqref{Matrix M}. The results are
~\\
\begin{empheq}{alignat=7}
	\label{O1}
	\Omega_E^{1,\alpha}&=\left\{
	\begin{array}{cc}
		{\displaystyle\left(\frac{1+\cos\theta_0}{2}\right)^{-\frac{1}{4}}} & \alpha E\leq-\frac{1}{2}
		\\\\
		{\displaystyle\left(\frac{1+\cos\theta_0}{2}\right)^{\frac{1}{4}}} & \alpha E\geq\frac{1}{2}
	\end{array}
	\right.\,,
\end{empheq}
~\\
\begin{empheq}{alignat=7}
	\label{O2}
	\Omega_E^{2,\alpha }&=\left\{
	\begin{array}{cc}
		{\displaystyle\left(\frac{1+\cos\theta_0}{2}\right)^{\frac{1}{4}}} & \alpha E\leq-\frac{1}{2}
		\\\\
		{\displaystyle\left(\frac{1+\cos\theta_0}{2}\right)^{-\frac{1}{4}}\left(\frac{\alpha E+\frac{1}{2}+\cos\theta_0}{\alpha E+\frac{3}{2}}\right)} & \alpha E\geq\frac{1}{2}
	\end{array}
	\right.\,,
\end{empheq}
~\\
\begin{empheq}{alignat=7}
	\label{O3}
	\Omega_E^{3,\alpha }&=\left\{
	\begin{array}{cc}
		{\displaystyle\left(\frac{1+\cos\theta_0}{2}\right)^{\frac{3}{4}}} & \alpha E\leq-\frac{1}{2}
		\\\\
		{\displaystyle\left(\frac{1+\cos\theta_0}{2}\right)^{-\frac{3}{4}}
		\left(\frac{\left(\alpha E+\frac{1}{2}\right)\left(\alpha E+\frac{3}{2}\right)-\frac{iM_2}{4M_1}\sin^2\theta_0}{\left(\alpha E+\frac{1}{2}\right)\left(\alpha E+\frac{3}{2}\right)}\right)} & \alpha E\geq\frac{1}{2}
	\end{array}
	\right.\,.
\end{empheq}
~\\
Here we have already taken the $\epsilon\to0$ and $R\to\infty$ limits. As with the bosonic determinants, it is convenient to combine the charge conjugate fields as
\begin{empheq}{alignat=7}
	\label{Om1}
	\ln\bm\Omega^1_E&\equiv\frac{1}{2}\left(\ln\Omega_{E}^{1,+}+\ln\Omega_{E}^{1,-}\right)&&=0\,,
	\\
	\label{Om2}
	\ln\bm\Omega^2_E&\equiv\frac{1}{2}\left(\ln\Omega_{E}^{2,+}+\ln\Omega_{E}^{2,-}\right)&&=\frac{1}{2}\ln\left(\frac{|E|+\frac12+ \cos\theta_0 }{|E|+\frac32}\right)\,,
	\\
	\label{Om3}
	\ln\bm\Omega^3_E&\equiv\frac{1}{2}\left(\ln\Omega_{E}^{3,+}+\ln\Omega_{E}^{3,-}\right)&&=\frac{1}{2}\ln\left(\frac{\left(|E|+\frac{1}{2}\right)\left(|E|+\frac{3}{2}\right)-\frac{iM_2}{4M_1}\sin^2\theta_0}{\left(|E|+\frac{1}{2}\right)\left( |E|+\frac{3}{2}\right)}\right)\,.
\end{empheq}

Notice that these expressions do not depend on the boundary conditions $N_1$ and $N_2$ at $\sigma=R$, and that only the determinants for the \emph{massless} fermions $\Omega^{3,\alpha}$ depend on the choice of boundary conditions $M_1$ and $M_2$ at $\sigma=0$. Moreover, for $M_2=iM_1$ the roots of the polynomial in the numerator of \eqref{Om3} drastically simplify, leading to the nice factorization
\begin{empheq}{alignat=7}\label{factorization}
	\left( | E|+\frac{1}{2}\right)\left( | E|+\frac{3}{2}\right)+\frac{1}{4}\sin^2\theta_0&=\left(|E|+1-\frac{\cos\theta_0}{2}\right)\left(|E|+1+\frac{\cos\theta_0}{2}\right)\,.
\end{empheq}
Of course, for general $M_1$ and $M_2$ a similar factorization is still possible, but the roots are more cumbersome. Ultimately, this is our main \emph{empirical} reason for this choice of boundary conditions; the correct $\theta_0$ dependence.

\subsection{Final result}
We are now ready to sum the ratios of determinants over the Fourier modes $E$. Given that equations \eqref{Om4}-\eqref{Om6} and \eqref{Om1}-\eqref{Om3} are  symmetric under $E\to-E$, we perform the summations with a symmetric cutoff $\Lambda$, namely,
\begin{empheq}{alignat=7}
	\sum_{E\,\in\,\mathds{Z}}\ln\bm\Omega_E&\longrightarrow \ln\bm\Omega_0+2\sum_{E=1}^{\Lambda}\ln\bm\Omega_E\,,
	&\qquad\textrm{for bosons}\,,
	\\
	\sum_{E\,\in\,\mathds{Z}+\frac{1}{2}}\ln\bm\Omega_E&\longrightarrow 2\sum_{E=0}^{\Lambda}\ln\bm\Omega_{E+\frac12}\,,
	&\qquad\textrm{for fermions}\,.
\end{empheq}
As argued in \cite{Faraggi:2016ekd}  in the context of type IIB strings, for $\Lambda\rightarrow\infty$ this coincides with a supersymmetric regularization scheme. Using this prescription, the sums for each charge conjugate pair of operators give
\begin{empheq}{alignat=7}
	\ln\bm\Omega^4_0+2\sum_{E=1}^{\Lambda}\ln\bm\Omega^4_E&=0\,,
	\\
	\ln\bm\Omega^5_0+2\sum_{E=1}^{\Lambda}\ln\bm\Omega^5_E&=\ln\left(\frac{\Gamma\left(\Lambda+2+\cos\theta_0\right)}{\Gamma\left(\Lambda+3\right)\Gamma\left(1+\cos\theta_0\right)}\right)-\frac{1}{2}\ln\left(\frac{1+\cos\theta_0}{2}\right)\,,
	\\
	\ln\bm\Omega^6_0+2\sum_{E=1}^{\Lambda}\ln\bm\Omega^6_E&=\ln\left(\frac{\Gamma\left(\Lambda+1+\frac{1+\cos\theta_0}{2}\right)}{\Gamma\left(\Lambda+2\right)\Gamma\left(\frac{1+\cos\theta_0}{2}\right)}\right)-\frac{1}{2}\ln\left(\frac{1+\cos\theta_0}{2}\right)\,,
	\\
	2\sum_{E=0}^{\Lambda}\ln\bm\Omega^1_{E+\frac12}&=0\,,
	\\
	2\sum_{E=0}^{\Lambda}\ln\bm\Omega^2_{E+\frac12}&=\ln\left(\frac{\Gamma\left(\Lambda+2+\cos\theta_0\right)}{\Gamma\left(\Lambda+3\right)\Gamma\left(1+\cos\theta_0\right)}\right)\,,
	\\
	2\sum_{E=0}^{\Lambda}\ln\bm\Omega^3_{E+\frac12}&=\ln\left(\frac{\Gamma\left(\Lambda+2+\frac{1+\cos\theta_0}{2}\right)\Gamma\left(\Lambda+3-\frac{1+\cos\theta_0}{2}\right)}{\Gamma\left(\Lambda+3\right)\Gamma\left(\Lambda+2\right)\Gamma\left(\frac{3+\cos\theta_0}{2}\right)\Gamma\left(\frac{3-\cos\theta_0}{2}\right)}\right)\,.
\end{empheq}
For the massless fermions $\bm\Omega^3_E$ we have set the boundary condition $M_2=iM_1$, although one can easily compute the sums for arbitrary $M_1$ and $M_2$. Combining the full spectrum in \eqref{eq6} with the correct multiplicities and taking into account the IR anomaly \eqref{IR anomaly},
\begin{empheq}{alignat=7}\label{Full sum}
	\Delta\Gamma^{\text{1-loop}}_{\text{effective}}&=\frac{1}{2}\Bigg[\sum_{E\,\in\mathds{Z}}\left(2\ln\bm\Omega^4_E+2\ln\bm\Omega^5_E+4\ln\bm\Omega^6_E\right)- \sum_{E\,\in\,\mathds{Z}+\frac{1}{2}}\left(2\ln\bm\Omega^1_E+4\ln\bm\Omega^2_E+2\ln\bm\Omega^3_E\right)\Bigg]+\Gamma_{\infty}\,,
\end{empheq}
we get
\begin{empheq}[box=\fbox]{alignat=7}
	e^{-\Delta\Gamma^{\text{1-loop}}_{\text{effective}}}&=\frac{\Gamma\left(\frac{1+\cos\theta_0}{2}\right)}{\Gamma\left(\frac{3-\cos\theta_0}{2}\right)\Gamma\left(1+\cos\theta_0\right)}\,.
\end{empheq}
Here we have already taken the $\Lambda\to\infty$ limit since the result is divergence-free. As advertised, this agrees with the field theory prediction \eqref{WLth}.
\section{Discussion}\label{Sec:discussion}
Having shown that the String Theory answer precisely matches the field theory result, we proceed to discuss in some detail the mechanisms for the cancellation of potential divergences. What we see in precision holography situations such as this one is a subtle balance between the effective number of bosonic and fermionic degrees of freedom. By effective we mean those modes in the spectrum of fluctuations that are not identical for two different string configurations.

When summing over Fourier modes, linear $\Lambda$ divergences are controlled by the $|E|\to\infty$ behavior of $\ln\bm\Omega_E$. As we can see from equations \eqref{Om4}-\eqref{Om6} and \eqref{Om1}-\eqref{Om3}, in all cases $\ln\bm\Omega_E\to0$, immediately implying the cancellation of linear divergences. Notice that the pairing of charge conjugate contributions into $\ln\bm\Omega_E=\frac{1}{2}\left(\ln\Omega_E^++\ln\Omega_E^-\right)$, as done above, is equivalent to combining $\ln\Omega^{\alpha}_E+\ln\Omega^{\alpha}_{-E}$ within each operator.

%
%
%

Regarding logarithmic divergences, the generic sums of the kind encountered here behave as
\begin{empheq}{alignat=7}
	\sum_{E}^{\Lambda}\ln\left(\frac{E^2+uE+s}{E^2+vE+t}\right)
	&=(u-v)\ln\Lambda+O(\Lambda^0)\,,
	&\qquad
	\sum_{E}^{\Lambda}\ln\left(\frac{E+u}{E+v}\right)
	&=(u-v)\ln\Lambda+O(\Lambda^0)\,,
\end{empheq}
where $u, v, s, t$ are real numbers. The fact that the fermionic sums are over half-integers does not affect this structure. In our case, except for the massless fermionic determinant, all of the sums are linear in $E$. Taking a closer look at \eqref{Full sum} we find that
\begin{empheq}{alignat=7}
	\nonumber
 	&\sum_{E}^{\Lambda}\left(2\ln\bm\Omega^4_E+2\ln\bm\Omega^5_E+4\ln\bm\Omega^6_E-2\ln\bm\Omega^1_E-4\ln\bm\Omega^2_E-2\ln\bm\Omega^3_E\right)=
 	\\
 	&\left[2\times0+2\times1+4\times\frac{1}{2}-2\times0-4\times1-2\times0\right]\left(\cos\theta_0-1\right)\ln\Lambda+O(\Lambda^0)=O(\Lambda^0)\,,
\end{empheq}
verifying that the total logarithmic divergence cancels indeed. We emphasize that, given the different multiplicities in the spectrum, it is crucial that the massless fermionic modes $\bm\Omega^3_E$ do not contribute to the divergence (since $u=v$ in \eqref{Om3}), regardless of the boundary conditions, and that the bosonic modes $\bm\Omega^6_E$ enter with a relative factor of $\frac{1}{2}$. At the risk of being repetitive, we compare this result with the analogous cancellation in type IIB described in \cite{Faraggi:2016ekd}. Recall that in that case the non-trivial bosonic contributions came from a pair of charged fields and a neutral triplet (denoted by 5,6 and 7,8,9 in \cite{Faraggi:2016ekd}), while all the fermionic modes had $\mathcal{O}^{2,\alpha}$ as their fluctuation operator. The potentially divergent piece of the one-loop effective action then took the form
\begin{empheq}{alignat=7}
 	\left[3\times0+2\times1+3\times2-8\times1\right]\left(\cos\theta_0-1\right)\ln\Lambda+O(\Lambda^0)&=O(\Lambda^0)\,.
\end{empheq}
Even though these cancellations are expected from general principles of string perturbation theory, it is satisfying to see the inner workings case by case. In the language of zeta-function regularization, such logarithmic divergences were explicitly discussed in \cite{Aguilera-Damia:2018twq,Aguilera-Damia:2018bam} and shown to be proportional to $\Delta\zeta(0)$, thus the association with the effective number of degrees of freedom  as  seen in the Gel'fand-Yaglom approach.

Finally, we comment on the boundary conditions for the fermionic modes. On the one hand, as seen in \eqref{O1}-\eqref{O3}, all the determinants turn out to be independent of the choice of boundary conditions $N_1$ and $N_2$ at $\sigma=R$. This is in agreement with the general expectation that, when putting a system in a finite box of length $R$, many of the details of the spectral density of eigenvalues are lost in the $R\to\infty$ limit. On the other hand, we found that only the massless fields are sensitive to the boundary conditions $M_1$ and $M_2$ at $\sigma=\epsilon$. This is related to the fact that, for massive fields, regularity at $\sigma=0$ discards half of the eigenfunctions, whereas for massless modes all eigenfunctions are regular and an additional condition needs to be imposed \cite{Medina-Rincon:2019bcc}. Furthermore, in order to get a precise match with the field theory prediction, we had to choose $M_2=iM_1$ in \eqref{Matrix M}, which, taking into account the relation \eqref{conjugate bc} for the charge conjugate operators, corresponds to setting
\begin{empheq}{alignat=7}
	\psi_1(\epsilon)+i\alpha\psi_2(\epsilon)&=0\,.
\end{empheq}
Written in a basis-independent way, this is equivalent to 
\begin{empheq}{alignat=7}
	\Pi_{\alpha}\psi(\epsilon)&=0\,,
	&\qquad
	\Pi_{\alpha}&=\frac{1}{2}\left(1+i\alpha n_{\mu}\gamma^{\mu}\gamma^*\right)\,.
\end{empheq}
The projectors $\Pi_{\pm}$ are precisely the ones used in \cite{Medina-Rincon:2019bcc}. However, contrary to the analysis of \cite{Medina-Rincon:2019bcc}, in our approach the boundary conditions for charge conjugate fermionic fields involve opposite projections. It would be interesting to understand the origin of this discrepancy.

\section{Conclusions}\label{Sec:Conclusions}
 
In this brief note, we have computed the one-loop effective action for the latitude string in AdS$_4\times \mathbb{CP}^3$ finding perfect agreement with the localization result in field theory.  We have further shown explicitly how divergences, known to be present in various other contexts, are cancelled in this case. It is interesting to note that the mechanism for cancellations is different in details from the one arising in the analogous context of holographic Wilson loops in Type IIB string theory on AdS$_5\times S^5$.  This understanding and control of the potential divergences is a necessary condition in the analysis of precision holography. 

A similar holographic computation with precise agreement with field theory has recently been reported in \cite{Medina-Rincon:2019bcc}, where the method of phase shifts was employed. Our results in this note elucidate the compatibility of the Gel'fand-Yaglom method to that of phase shifts for this problem. These are currently some of the most popular methods in the tool box required for precision holography with Wilson loops.  Although we hope to discuss such equivalence somewhere else, it is clear that the methods deal with similar difficulties, such as boundary conditions, in their own idiosyncratic ways. 

It is certain that progress has been made in the field of precision holography with Wilson loops. Various one-loop computations can now be clearly sketched and compared among themselves. There are, nevertheless, a number of important questions that would be useful to clarify. We leave a number of interesting question for future work. Most pressing in our view is a rigorous proof of the equivalence between the phase shifts and Gel'fand-Yaglom methods and their connection to the zeta-function regularization. 

A particularly interesting avenue to test many of these ideas is the problem of the $k$-wound Wilson loops in ${\cal N}=4$ SYM and in the ABJM theory; in both cases, the field theory answers are known exactly and can be readily extrapolated to the regime where a comparison with string theory is appropriate. Indeed, a number of attempts has been taken with the goal of matching the field theory result with the effective action of string configurations without achieving an exact match \cite{Kruczenski:2008zk,Bergamin:2015vxa,Forini:2017whz}. We hope to report on this fascinating problem in the near future.  

Considering the success of precision holography for Wilson loops dual to string configurations, it may be time to revisit Wilson loops whose dual are branes. There was one attempt for the $\frac{1}{2}$-BPS Wilson loop in ${\cal N}=4$  in the totally symmetric representation \cite{Buchbinder:2014nia}. The dual of this system is a D3 brane and in this case, the fluctuations were completely described in \cite{Faraggi:2011bb}. In addition, it would be worth revisiting the one-loop effective action of the Wilson loop in the anti-symmetric representation whose dual is a D5 brane \cite{Faraggi:2011ge}. A summary of the situation for higher representations Wilson loops was given in \cite{Faraggi:2014tna}, where a number of discrepancies was noted. 

\section*{Acknowledgments}
We are thankful to Luca Griguolo and Itamar Yaakov for various clarifying discussions. MD, RdLA and AF are grateful to the ICTP for warm hospitality during the final stages of this project. MD is supported by the Center for European Studies of the University of Michigan. This material is based upon work supported by the National Science Foundation Graduate Research Fellowship under Grant No. DGE 1256260. MD and LAPZ are partially supported by the US Department of Energy under Grant No. DE-SC0007859. GAS's work is supported by UNLP and CONICET grants X791, PIP 2017-1109 and PUE B\'usqueda de nueva F\'isica.

\bibliographystyle{JHEP}
\bibliography{WL-ABJM}
\end{document}